\documentclass[pra,twocolumn,showpacs, amsmath,amssymb,aps]{revtex4-1}
\usepackage{graphicx}
\usepackage{bm}

\begin{document}
\newcommand{\balpha}{\mbox{\boldmath$\alpha$}}

\title{Spaser quenching by off-resonant plasmon modes}

\author{L. S. Petrosyan$^{1,2}$ and T. V. Shahbazyan$^1$} 
\affiliation{
$^1$Department of Physics, Jackson State University, Jackson, Mississippi
39217 USA\\
$^2$Institute for Mathematics and High Technology, Russian-Armenian State University, 123 Hovsep Emin Street, Yerevan  0051, Armenia
}


\begin{abstract}
We study the effect of off-resonant plasmon modes on spaser threshold in nanoparticle-based spasers. We develop an analytical semiclassical model and derive spaser threshold condition accounting for gain coupling to higher-order plasmons. We show that such a coupling originates from inhomogeneity of gain distribution near the metal surface and leads to an upward shift of spaser frequency and population inversion threshold. This effect is similar, albeit significantly weaker, to quenching of plasmon-enhanced fluorescence near metal nanostructures due to excitation of off-resonant modes with wide spectral band. We also show that spaser quenching is suppressed for high gain concentrations and establish a simple criterion for quenching onset, which we support by numerical calculations for spherical geometry.
\end{abstract}

\pacs{78.67.Bf, 73.20.Mf, 33.20.Fb, 33.50.-j}

\maketitle

\section{Introduction}
\label{sec:intro}
The prediction of plasmonic laser (spaser) \cite{bergman-prl03,stockman-natphot08,stockman-jo10} and its  experimental realization in various systems \cite{noginov-nature09,zhang-nature09,zheludev-oe09,zhang-natmat10,ning-prb12,gwo-science12,odom-natnano13,shalaev-nl13,gwo-nl14,zhang-natnano14,odom-natnano15} have been among the highlights of the rapidly developing field of plasmonics during the  past decade \cite{stockman-review}. First reported in gold nanoparticles (NPs) coated by dye-doped silica shells \cite{noginov-nature09}, spaser action was observed in hybrid plasmonic waveguides \cite{zhang-nature09}, semiconductor quantum dots on metal film \cite{zheludev-oe09,gwo-nl14},  plasmonic nanocavities and nanocavity arrays \cite{zhang-natmat10,ning-prb12,gwo-science12,odom-natnano13,zhang-natnano14,odom-natnano15},    metallic NPs and nanorods \cite{noginov-nature09,shalaev-nl13}, and more recently, carbon-based structures \cite{apalkov-light14,premaratne-acsnano14} and hyperbolic materials \cite{pustovit-prb16-2}. Small  spaser size  well below the diffraction limit gives rise to a wealth of promising applications  \cite{stockman-aop17}.

\begin{figure}[bt]
 \centering
 \includegraphics[width=1\columnwidth]{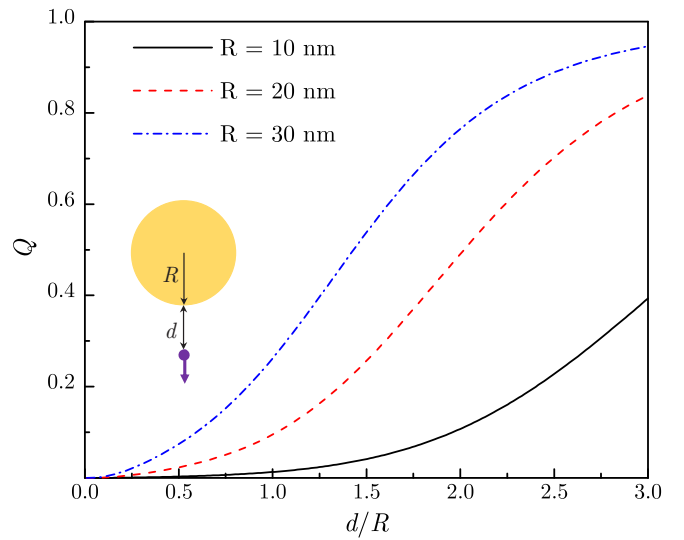}
 \caption{\label{fig1} Fluorescence quantum efficiency for a QE near spherical Au NP is shown vs. QE-NP distance  for several NP sizes.
  }
 \vspace{-5 mm}
  \end{figure}

The spaser feedback mechanism is based on  energy transfer (ET) between quantum emitters (QEs), constituting gain medium, and  resonant plasmon mode. Even though a metal nanostructure possesses  discrete spectrum of localized plasmon modes, e.g., characterized by angular momentum $l$ for spherical systems,  the QE coupling to \textit{off-resonant} modes well separated in frequency from QE  (and from resonant mode)  is usually considered sufficiently weak and, hence, neglected \cite{wegener-oe08,stockman-jo10,klar-bjn13,li-prb13,lisyansky-oe13,bordo-pra13}. However, while this is a good approximation for high-quality cavity modes, the plasmon resonances are characterized by much broader bands due to large Ohmic losses in metal, so that a significant fraction of  excited QE energy is transferred to off-resonant modes, especially for small QE distances to the metal surface and, correspondingly, large QE-plasmon coupling \cite{nitzan-jcp81,ruppin-jcp82,pustovit-prl09,pustovit-prb10}.  In plasmon-enhanced fluorescence spectroscopy, such processes lead to distance-dependent radiation quenching \cite{feldmann-prl02,lakowicz-jf02,artemyev-nl02,gueroui-prl04,feldmann-nl05,klimov-jacs06,strose-jacs06,mertens-nl06,pompa-naturenanotech06,novotny-prl06,sandoghdar-prl06,novotny-oe07,sandoghdar-nl07,lakowicz-jacs07,chen-nl07,lakowicz-nl07,halas-nl07,feldmann-nl08,halas-acsnano09,ming-nl09,kinkhabwala-naturephot09,viste-acsnano10,lakowicz-jacs10,munechika-nl10,ming-nl11,ratchford-nl11,raino-acsnano11}, characterized by quantum efficiency $Q=\Gamma_{r}/(\Gamma_{r}+\Gamma_{nr})$, where $\Gamma_{r}$ is plasmon-enhanced radiative decay rate and $\Gamma_{nr}$ is nonradiative decay rate due to QE coupling to higher-order modes (see below for detail). To illustrate the role of off-resonant modes in fluorescence quenching, in Fig.~\ref{fig1} we plot quantum efficiency of a radiating dipole at distance $d$ from the surface of a spherical NP of radius $R$. Even though  the QE radiative rate $\Gamma_{r}$ is enhanced due to coupling to resonant dipole plasmon mode, the decay a QE  into off-resonant dark modes, characterized by the rate $\Gamma_{nr}$, becomes the dominant process as  $d$ decreases, and so $Q$ is significantly reduced at distances $d\sim R$.

In spasers, the effect of gain coupling to off-resonant  modes is twofold. First, the QE coupling to higher-order plasmons (with higher frequencies) should lead to an upward shift of spaser frequency; second, the ET from QEs to off-resonant modes can interfere with the feedback mechanism, resulting in higher population inversion threshold (we assume that higher-order modes are sufficiently separated in frequency from the resonant mode so that no instabilities arise \cite{hess-prb12,hess-science13}). Both effects have an increasingly negative impact  on spaser action as the (average) distance between QEs and the metal surface is reduced, which raises the issue of \textit{spaser quenching}  for substantially close gain-metal proximity.

In recent work \cite{pustovit-prb16}, we carried out a numerical study of the role of  gain coupling to off-resonant modes as well as of direct dipole coupling between QEs in small NP-based spasers. Due to numerical challenges, our simulations were restricted to relatively small (5 nm radius) NPs with thin (up to 2.5 nm) dye-doped  dielectric shells and relatively low (up to 1000)  numbers of QEs with dipole moments oriented normally to the NP surface. In such systems, the direct coupling between gain molecules is maximal, which leads to random Coulomb shifts of molecule excitation energies  and, hence, to dephasing \cite{friedberg-pra74,stockman-prl97}. For small systems, the spasing eigenstates were found via exact numerical diagonalization \cite{pustovit-prb16}; however, for larger systems with realistic random dipole orientations, direct numerical determination of many-body eigenstates of interacting QEs is not feasible.

At the same time, for large ensembles of randomly-oriented QEs, the ensemble-averaged dipole coupling between individual QEs vanishes, while weak fluctuations of QEs' excitation energies do not significantly affect the collective system eigenstates strongly coupled to radiation \cite{shahbazyan-prb00}. In this case, a major source of dephasing in NP-based spasers is the ET between gain and off-resonant plasmon modes which is largely insensitive to QE dipole orientations. Although the ET rate between a QE and off-resonant modes is normally significantly lower than between QE and resonant mode, the number of excited modes increases exponentially as the QE distance to NP surface is reduced \cite{nitzan-jcp81,ruppin-jcp82,pustovit-prl09,pustovit-prb10}, which leads to significant fluorescence quenching for distances below NP radius \cite{feldmann-prl02,lakowicz-jf02,artemyev-nl02,gueroui-prl04,feldmann-nl05,klimov-jacs06,strose-jacs06,mertens-nl06,pompa-naturenanotech06,novotny-prl06,sandoghdar-prl06,novotny-oe07,sandoghdar-nl07,lakowicz-jacs07,chen-nl07,lakowicz-nl07,halas-nl07,feldmann-nl08,halas-acsnano09,ming-nl09,kinkhabwala-naturephot09,viste-acsnano10,lakowicz-jacs10,munechika-nl10,ming-nl11,ratchford-nl11,raino-acsnano11} (see Fig.~\ref{fig1}). The main effect of gain coupling to off-resonant modes on spaser action is the disruption of the gain-plasmon feedback, and, hence, the increase of the spasing threshold (spaser quenching). According to our numerical simulations  \cite{pustovit-prb16}, the effect of quenching on spaser action is much weaker than on single-molecule fluorescence. However, no analytical model for spaser quenching  and, importantly, no spaser condition accounting for off-resonant modes has  so far been suggested. The goal of this paper is to provide such a model.

Specifically, we develop an analytical model for  plasmonic systems with gain that includes gain coupling to off-resonant modes.  We show that the interplay between such coupling   and inhomogeneity of gain distribution near the metal surface leads to an upward shift of spaser frequency and increases population inversion threshold. At the same time, we demonstrate that, with increasing gain concentration, the role of off-resonant modes  is reduced  and their overall effect on spaser action is much weaker than on single-molecule fluorescence. For NP-based spasers, we obtain an explicit spaser condition that accounts for off-resonant modes, and derive a simple criterion for spaser quenching onset in terms of system parameters, which we support by numerical calculations.

The paper is organized as follows. In Sec.~\ref{sec2}, we set up Maxwell-Bloch equations for pumped QEs interacting with a composite spherical NP closely following our previous work \cite{pustovit-prb16}. In Sec. \ref{sec3}, we introduce the system collective modes and show that the effect of off-resonant modes on spaser feedback mechanism hinges on inhomogeneity of gain distribution near the NP surface. In Sec.~\ref{sec4}, we derive the spaser condition that accounts for off-resonant modes and provide a simple criterion for quenching onset, which we supplement by numerical calculations. In Sec.~\ref{sec5}, we discuss the approximations made and summarize our findings. 

\section{Pumped quantum emitters interacting with a metal nanoparticle}
\label{sec2}

In this section, we outline a semiclassical approach for metal NP-based spasers based on Maxwell-Bloch equations \cite{bergman-prl03,stockman-natphot08,stockman-jo10} following closely the notations of our previous paper \cite{pustovit-prb16}. We consider a thin layer of $M$ QEs randomly distributed on top of a spherical core-shell NP with metal core of radius  $R$ and a dielectric shell of uniform thickness $d$. Within the semiclassical approach, electromagnetic fields are treated classically, while QEs are described by  pumped two-level systems  located at $\textbf{r}_{j}$ with excitation frequency $\omega_{12}$ between energy levels 1 and 2. Each QE is  characterized by  polarization $\rho_{j} \equiv \rho_{12}^{(j)}$ and occupation $n_{j} \equiv\rho_{22}^{(j)}-\rho_{11}^{(j)}$, where $\rho_{ab}^{(j)}$ ($a,b=1,2$) is the density matrix for $j$th QE. The ensemble population inversion is $N=\sum_{j}n_{j} $. In the rotating wave approximation, the steady-state dynamics of QEs coupled to  alternating electric field $\bm{\mathcal{\cal E}}(\bm{r})e^{-i\omega t}$ is described by the standard Maxwell-Bloch equations 
\begin{align}
\label{maxwell-bloch}
& \left(\omega - \omega_{21} +i/\tau_{2}\right) \rho_{j} =\frac{\mu}{\hbar} \, n_{j} \,\bm{e}_{j}\!\cdot\! \bm{\mathcal{\cal E}}(\bm{r}_{j}),
\\
& n_{j} -\bar{n}  =-  \frac{4\mu\tau_{1}}{\hbar} \,   \text{Im} \! \left[\rho_{j} \,\bm{e}_{j}\!\cdot\! \bm{\mathcal{\cal E}}(\bm{r}_{j})\right],
\nonumber
\end{align}
where $\tau_2$ and  $\tau_1$ are  time constants characterizing polarization and population relaxation, $\mu$ and  $\bm{e}_{j}$ are, respectively, the QE dipole matrix element and orientation, and $\bar{n}$ is the average population inversion per QE due to the pump.  The  local field $\bm{\mathcal{\cal E}}(\bm{r}_{j})$ is generated by all QEs with dipole moments $\bm{p}_{j}=\mu \bm{e}_{j}\rho_{j}$ and, within the semiclassical approach, has the form
\begin{equation}
\label{electric}
\bm{\mathcal{\cal E}}(\bm{r}_{j} ) =\frac{4\pi\omega^{2}}{c^{2}}\sum_{k} \bar{\textbf{G}}(\omega;\bm{r}_{j},\bm{r}_{k}) \!\cdot\! \bm{p}_{k},
\end{equation}
where $\bar{\textbf{G}}(\omega;\bm{r},\bm{r}')$ is the electromagnetic Green dyadic in the presence of metal nanostructure and $c$ is the speed of light. 
Using Eq.~(\ref{electric}) to eliminate the electric field, the system Eq.~(\ref{maxwell-bloch}) takes the form
\begin{align}
\label{maxwell-bloch2}
&\sum_{k=1}^{M}\left [\left (\omega - \omega_{21} +\frac{i}{\tau_{2}}\right )\delta_{jk}-\frac{\mu^{2}}{\hbar} \, n_{j}D_{jk}\right ]\rho_{j}=0,
\nonumber\\
&n_{j}-\bar{n} +\frac{4\tau_{1}\mu^{2}}{\hbar} \, \text{Im}\sum_{k=1}^{M}\left( \rho_{j}^{*}D_{jk}\rho_{j}\right)=0,
\end{align}
where $\delta_{jk}$ is Kronecker symbol and $D_{jk}(\omega)$ is a frequency-dependent coupling matrix in position space,
\begin{equation}
\label{matrix}
D_{jk}(\omega)=\dfrac{4\pi\omega^{2}}{c^{2}}\,\bm{e}_{j} \!\cdot \!  \bar{\textbf{G}}(\omega;\bm{r}_{j},\bm{r}_{k}) \! \cdot \! \bm{e}_{k}.
\end{equation}
For small system sizes well below the radiation wavelength, the Green dyadic can be replaced by its near-field limit, and  the coupling matrix Eq.~(\ref{matrix}) represents a sum of direct and plasmon terms, $D_{jk}=D_{jk}^{0}+D_{jk}^{p}$, which, for spherical geometry,  are given by \cite{pustovit-prl09,pustovit-prb10}
\begin{align}
\label{D-simple}
&D_{jk}^{0}=-\sum_{lm}\left [\psi_{lm}^{(j)}\chi_{lm}^{(k)\ast} \theta_{jk}  + \chi_{lm}^{(j)}\psi_{lm}^{(k)\ast} \theta_{kj}\right ],
\nonumber\\
&D_{jk}^{p}=\sum_{lm} \alpha_{l}\psi_{lm}^{(j)}\psi_{lm}^{(k)\ast} ,
\end{align}
where $l$ and $m$ are the polar and azimuthal numbers, respectively, and  $\theta_{jk}\equiv\theta(r_{j}-r_{k})$ is the step-function. Here, $\alpha_{l}(\omega)$ is   $l$-pole polarizability for a spherical NP in a medium with dielectric constant $\varepsilon_{d}$,
\begin{equation}
\label{alpha}
\alpha_{l}(\omega)=\frac{R^{2l+1}(\varepsilon-\varepsilon_{d})}{\varepsilon+(1+l^{-1})\varepsilon_{d}},
\end{equation}
where $\varepsilon(\omega)$ is the metal dielectric function. The basis functions are given by
\begin{align}
\chi_{lm}^{(j)}=C_{l} \bm{e}_{j} \! \cdot \! {\bm \nabla}_{j} \! \left [r_{j}^{l}Y_{lm}(\hat{\bm r}_{j})\right ],
~
\psi_{lm}^{(j)}=C_{l} \bm{e}_{j} \! \cdot \! {\bm \nabla}_{j} \! \left [\frac{Y_{lm}(\hat{\bm r}_{j})}{r_{j}^{l+1}}\right ]\!  ,
\end{align}
where $C_{l}= \sqrt{4\pi/(2l+1)}$ is normalization coefficient and $Y_{lm}(\hat{\bm r})$ are the spherical harmonics.  The basis functions satisfy  orthogonality relations 
\begin{align}
\label{chipsi}
&\langle \chi_{lm}^{(j)\ast}\chi_{l'm'}^{(j)}\rangle=\frac{l}{3}r_{j}^{2l-2}\delta_{ll'}\delta_{mm'},\\
&\langle \psi_{lm}^{(j)\ast}\psi_{l'm'}^{(j)}\rangle=\frac{1}{3}\frac{l+1}{r_{j}^{2l+4}}\delta_{ll'}\delta_{mm'},
~~~\langle \chi_{lm}^{(j)\ast}\psi_{l'm'}^{(j)}\rangle=0,
\nonumber
\end{align}
where brackets stand for angular and orientational averaging. Note that,   for QEs with random dipole orientations and uniformly distributed in the shell, the direct term $D_{jk}^{0}$ in system Eq.~(\ref{D-simple}) \textit{vanishes} on average, so we keep only the plasmon term $D_{jk}^{p}$ in the following.


\section{Collective modes of quantum emitters and spaser condition}
\label{sec3}

Within semiclassical approach, the first (homogeneous) equation in the system Eq.~(\ref{maxwell-bloch2}) determines the spaser condition. We now make transformation from individual QE representation to collective mode representation by introducing collective polarizations as 
\begin{equation}
\rho_{\lambda}=\sum_{j=1}^{M}\psi_{\lambda}^{(j)\ast}\rho_{j},
\end{equation}
where $\lambda=(lm)$ is the collective mode composite index. Keeping only the plasmon term $D_{jk}^{p}$ in the coupling matrix, multiplying the first  equation by $\psi_{\lambda}^{(j)\ast}$ and summing up over $j$, the system Eq.~(\ref{maxwell-bloch2}) takes the form
\begin{align}
\label{mb-coll}
&\left (\omega-\omega_{21}+ i/\tau_{2}\right )\rho_{\lambda} - \sum_{\lambda'}S_{\lambda\lambda'}\alpha_{\lambda'}\rho_{\lambda'}=0,
\nonumber\\
&N-\bar{N}  +  \frac{4\tau_1\mu^{2}}{\hbar}\sum_{\lambda} \alpha''_{\lambda}(\omega)\left |\rho_{\lambda}\right|^{2}=0,
\end{align}
where $N=\sum_{j}n_{j}$ is gain population inversion,  $\bar{N}=\bar{n}M$ is that due to the pump, and
\begin{equation}
S_{\lambda\lambda'}=\frac{\mu^{2}}{\hbar}\sum_{j=1}^{M}\psi_{\lambda}^{(j)\ast}n_{j}\psi_{\lambda'}^{(j)} 
\end{equation}
is the  mode coupling matrix.

\subsection*{Single mode approximation}

Let us perform angular and orientational averaging directly in the system Eq.~(\ref{mb-coll}). In the leading order in $1/M$, the coupling matrix $S$ can be replaced with its average,
\begin{equation}
\label{sl}
\langle S_{\lambda\lambda'}\rangle = s_{l}\delta_{ll'}\delta_{mm'},
~~
s_{l}=\frac{l+1}{3}\frac{\mu^{2}}{\hbar}\sum_{j}\frac{n_{j}}{r_{j}^{2l+4}},
\end{equation}
where we used relations Eqs.~(\ref{chipsi}), yielding the consistency condition for each  mode
\begin{equation}
\label{spasing}
\omega-\omega_{21}+  i/\tau_{2}= s_{l}\alpha_{l}(\omega).
\end{equation}
The real and imaginary parts of Eq.~(\ref{spasing}) determine, respectively, the spaser frequency and threshold population inversion:
\begin{equation}
\label{spasing2}
\omega-\omega_{21} =  s_{l}\alpha'_{l}(\omega),
\\
~~~~
 \tau_{2} s_{l}\alpha''_{l} (\omega) =1.
\end{equation}
By taking their ratio, the spasing frequency $\omega_{s}$ is obtained from simple equation
\begin{equation}
\label{frequency}
\tau_{2}(\omega-\omega_{21} )=\frac{\alpha'_{l}(\omega)}{\alpha''_{l}(\omega)}.
\end{equation}
Note that since Eqs.~(\ref{spasing2}) are independent of azimuthal number $m$, each $l$-mode is $(2l+1)$-fold degenerate.

Assume now that gain molecules are uniformly distributed in a thin layer at approximately equal distance $d$ from the metal  NP surface  (e.g., on top of dielectric shell), so that the averaged coupling  [Eq.~(\ref{sl})] takes the form
\begin{equation}
\label{sl2}
s_{l}=\frac{\mu^{2}}{\hbar }\frac{(l+1)N}{3(R+d)^{2l+4}},
\end{equation}
For QE frequency $\omega_{21}$ close to the \textit{l-pole} plasmon resonance frequency $\omega_{l}$, the NP polarizability can be expanded near the plasmon pole as
\begin{align}
\label{alpha1}
\alpha_{l}(\omega)= \frac{B_{l}}{\omega_{l}-\omega-i/\tau_{l}},
\end{align}
where $\tau_{l}=[\partial \varepsilon'(\omega_{l})/\partial \omega_{l}]/\varepsilon''(\omega_{l})$ is  plasmon lifetime and coefficient $B_{l}$ depends on NP shape and composition \cite{stockman-review}. For spherical NP, $B_{l}$ is obtained from Eq.~(\ref{alpha}) as
\begin{equation}
B_{l}=\frac{(2l+1)\varepsilon_{d}R^{2l+1}}{l\partial \varepsilon'(\omega_{l})/\partial \omega_{l}},
\end{equation}
where the factor $2l+1$ reflects the mode degeneracy. With NP polarization in the form of Eq.~(\ref{alpha1}),  we obtain from Eq.~(\ref{frequency})  the standard  spaser frequency   \cite{bergman-prl03,stockman-natphot08,stockman-jo10}
\begin{equation}
\label{frequency0}
\omega_{0}=\frac{\tau_{l} \omega_{l}+\tau_{2}\omega_{21}}{\tau_{l}+\tau_{2}},
\end{equation}
while the second equation in system Eqs.~(\ref{spasing2}) determines, for $|\omega_{l}-\omega_{21}|\tau_{l}\ll 1$, the  population inversion \textit{threshold} $N_{0}$,
\begin{equation}
\label{threshold0}
\frac{\mu^{2}\tau_{2}}{\hbar}  \frac{(2l+1)(l+1)}{3l \varepsilon''(\omega_{l})}  \frac{N_{0} R^{2l+1}}{(R+d)^{2l+4}}= 1.
\end{equation}
Note that  $N_{0}$ depends sensitively on the mode angular momentum. For $l=1$, we recover threshold population inversion for the dipole plasmon mode  \cite{bergman-prl03,stockman-natphot08,stockman-jo10},
\begin{equation}
\label{threshold1}
N_{0}=\frac{\hbar\varepsilon''(\omega_{l})R^{3}} {2\mu^{2}\tau_{2}}   \left (1+\frac{d}{R}\right )^{6}.
\end{equation}
However, for large angular momenta, the value of $N_{0}$  increases exponentially with $l$, implying that   feedback  via high-$l$ modes is ineffective. 

\section{Off-resonant modes and spaser quenching}
\label{sec4}

In this section, we incorporate,  within a semiclassical approach, the effect of higher-order plasmon modes on spaser action. While in the absence of gain, different plasmon modes  are orthogonal, the presence of QEs with random positions and orientations violates the underlying NP symmetry and leads to modes' coupling. For large number $M$ of randomly oriented QEs uniformly distributed around the NP, the spherical symmetry is preserved \textit{on average}, so that single-mode description is reasonably accurate, while corrections due to the modes' coupling are suppressed by a factor of $1/M$. However, for QEs located close to the NP surface, the coupling to off-resonant modes is strong, so that even weak inhomogeneity of QE distribution can lead to significant mode coupling effects.  Below we analyze the effect of off-resonant modes on spaser condition and establish a simple criterion, in terms of system parameters, for the validity of single-mode description.

\subsection{Spaser condition}

We assume that QE frequency $\omega_{21}$ is tuned to the dipole plasmon mode ($l=1$) frequency $\omega_{1}$, and incorporate the  effect of higher ($l>1$) off-resonant modes as follows. First, we separate out the resonant and higher-order modes in  the first equation of  system Eq.~(\ref{mb-coll})  by splitting it into two equations,
\begin{align}
\label{maxwell-bloch-p2}
& \Omega \rho_{1} - S_{11}\alpha_{1}\rho_{1} -\sum_{\lambda} S_{1\lambda}\alpha_{\lambda}\rho_{\lambda}=0,
\nonumber\\
& \Omega \rho_{\lambda} - S_{\lambda 1}\alpha_{1}\rho_{1} -\sum_{\lambda'} S_{\lambda\lambda'}\alpha_{\lambda'}\rho_{\lambda'}=0,
\end{align}
where we denoted $\Omega = \omega-\omega_{21}+i /\tau_{2} $, and  the indexes $\lambda$ and $\lambda'$ do \textit{not} include the resonant mode. In the first order in $1/M$, we include the coupling of resonant mode to off-resonant modes, but disregard off-resonant modes' coupling to each other. After replacing the matrix $S_{\lambda\lambda'}$ in the second equation by its average [Eq.~(\ref{sl})],  the polarization for off-resonant modes can be expressed via that for the resonance mode as
\begin{equation}
\rho_{\lambda}=\frac{S_{\lambda 1}\alpha_{1}}{\Omega-s_{\lambda}\alpha_{\lambda}}\,\rho_{1}.
\end{equation}
Then, eliminating  $\rho_{\lambda}$ from the first equation of system Eqs.~(\ref{maxwell-bloch-p2}), we obtain  the consistency condition [restoring indexes $(lm)$],
\begin{equation}
\label{spasing3}
\Omega \delta_{mm'}-\left (S_{1m,1m'}+\sum_{l_{1}m_{1}}
\frac{S_{1m,l_{1}m_{1}}\alpha_{l_{1}}S_{l_{1}m_{1}, 1m'}}{\Omega-s_{l_{1}}\alpha_{l_{1}}}\right )\alpha_{1}=0,
\end{equation}
where the sum  runs over  $l_{1}>1$ and $m_{1}$  in the interval $(-l_{1},l_{1})$, while $m,m'$ take values $(-1,0,1)$. Performing angular and orientational averaging in Eq.~(\ref{spasing3}) (see Appendix A), we finally arrive at  spaser condition:
\begin{equation}
\label{spasing5}
\omega-\omega_{21} +i/\tau_{2}-s_{1}\alpha_{1}(\omega)\left [1+f(\omega)\right ]=0,
\end{equation}
where the function 
\begin{equation}
\label{R}
f(\omega)=\frac{1}{5M}\sum_{l>1}\frac{(11l+7)s_{l}\alpha_{l}(\omega)}{\omega-\omega_{21} +i/\tau_{2}-s_{l}\alpha_{l}(\omega)}
\end{equation}
includes gain coupling to off-resonant plasmon modes. The new spaser condition Eq.~(\ref{spasing5}) is the central result of our paper, and below we estimate the quenching onset and present the results of   numerical calculations.

\subsection{Quenching onset}

In the absence of gain coupling to off-resonant modes ($f=0$), the solutions of Eq.~(\ref{R}) for  spaser frequency $\omega_{0}$ and threshold population $N_{0}$ are given, respectively, by Eqs.~(\ref{frequency0}) and (\ref{threshold0}). In the presence of such coupling, the corresponding solutions  $\omega$ and $N$ deviate from $\omega_{0}$ and $N_{0}$ by the amount depending on distance $d$ to the  NP surface. While for $d\gtrsim R$, the coefficients $s_{l}$, given by Eq.~(\ref{sl2}), change rapidly with $d$, for $d\ll R$ they are only weakly dependent on $d$, indicating that, in this case, the main contribution to $f$ comes  from  high-$l$ modes. To estimate the characteristic distance $d$ below which off-resonant modes become important, we note that for off-resonant modes we have $\tau_{2}s_{l}\alpha_{l}\ll 1$ and so  the last term in the denominator of Eq.~(\ref{R}) can be disregarded (this approximation is \textit{not} made in the numerical calculations below). Since the main contribution comes from high-$l$ terms,  we can replace $\alpha_{l}$ by $R^{2l+1}[\varepsilon(\omega)-\varepsilon_{d}]/[\varepsilon(\omega)+\varepsilon_{d}]$ [see Eq.~(\ref{alpha})] and write
\begin{equation}
f(\omega)=\frac{\mu^{2}}{15 \hbar } \, \frac{N}{M} \, \frac{\varepsilon(\omega)-\varepsilon_{d}}{\varepsilon(\omega)+\varepsilon_{d}} \, 
\frac{g}{\omega-\omega_{21} +i/\tau_{2}},
\end{equation}
where 

\begin{equation}
g=\sum_{l>1}(11l+7)(l+1) \frac{R^{2l+1}}{(R+d)^{2l+4}}.
\end{equation}
For $d/R\ll 1$, replacing the sum over $l$ by the integral, we obtain  $g\approx 11/4d^{3}$. For small deviations of $\omega$ from the plasmon frequency, i.e., $\varepsilon(\omega)\approx \varepsilon(\omega_{1})=-2\varepsilon_{d}$, and using that $(\omega-\omega_{21})\tau_{2}\ll 1$, we finally obtain
\begin{equation}
\label{R1}
|f| \approx \frac{\mu^{2}\tau_{2}}{2\hbar d^{3}} \frac{N}{M}.
\end{equation}
The onset of quenching corresponds to  $|f|\sim 1$. In the first order, replacing $N$ with $N_{0}=R^{3} \varepsilon''(\omega_{1})\hbar/2\mu^{2}\tau_{2}$  from Eq.~(\ref{threshold0}), we arrive at the estimate for onset value of $d$:
\begin{equation}
\label{estimate}
 d  \sim R \left [\frac{\varepsilon''(\omega_{1})}{4M}\right ]^{1/3},
\end{equation}
which decreases with increasing QE number $M$. As an example, for $M\sim 10^{3}$ and with $\varepsilon''(\omega_{1})\approx 2 $ for spherical Au NP, the high-$l$ modes are important for $d/R\lesssim 0.1$. 

\subsection{Numerical results}

Below we present the results of the numerical solution of the spaser condition [Eq.~(\ref{spasing5})], which  includes off-resonant modes, for spherical Au NP of radius $R$ and $M$ QEs randomly distributed on top of dielectric shell at distance $d$ from the metal surface  with   frequencies $\omega_{21}$ tuned to the dipole plasmon resonance frequency $\omega_{1}$. In all calculations, we used   experimental Au dielectric function \cite{christy} and included modes with angular momenta up to $l_{\rm max}=50$. Note that we excluded the region of very small distances dominated by quantum effects, which are beyond the scope of this paper \cite{pustovit-jcp12}.

 \begin{figure}[bt]
 \centering
 \includegraphics[width=1\columnwidth]{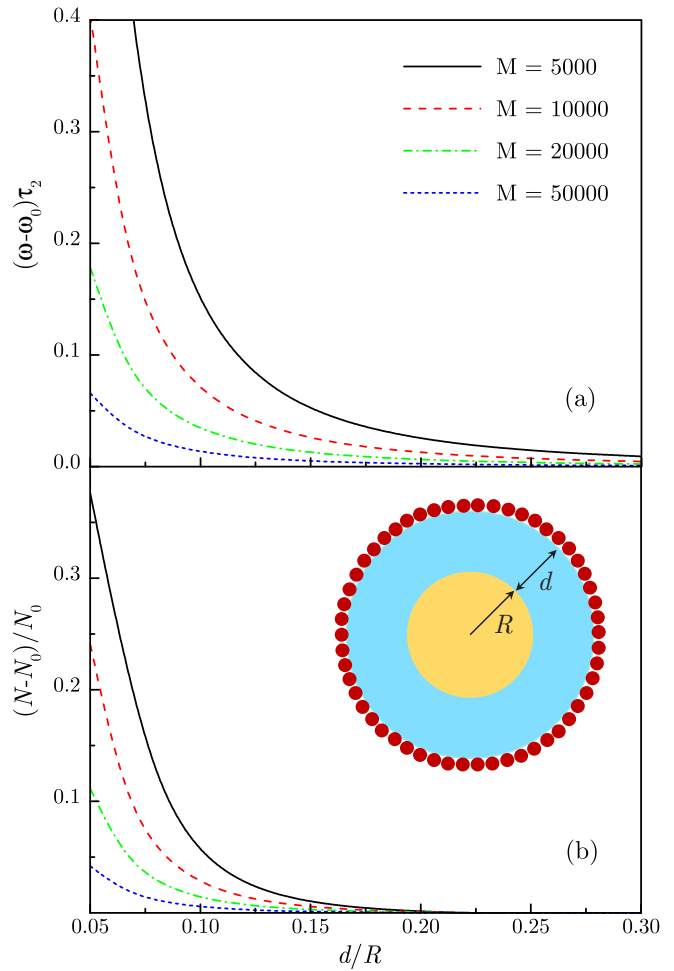}
 \caption{\label{fig2} (a) Spaser frequency shift and (b) relative  population inversion threshold shift are shown vs. shell thickness  for several QE ensemble sizes. Inset: Schematics of QEs distributed on top of composite NP.}
\vspace{-5 mm}
  \end{figure}

In Fig.~\ref{fig2}, we plot the spaser frequency $\omega$ and population inversion threshold $N$ vs. gain-NP distance $d$ (see inset)  obtained by solving Eq.~(\ref{spasing5}) for several ensemble sizes $M$. To highlight the role of off-resonant modes, we show the deviations of $\omega$ and $N$ from the values $\omega_{0}$ and $N_{0}$, respectively, corresponding to coupling only to resonant mode (i.e., $f=0$).  For small distances $d/R\ll 1$, the effect of off-resonant modes can be substantial depending on the ensemble size, consistent with our estimate [Eq.~(\ref{estimate})]. With decreasing $d$, the spaser frequency $\omega$ shifts upwards (high-order modes have larger frequencies), and so does the threshold $N$  to compensate  the energy leakage to off-resonant modes not participating in the feedback. At the same time, with increasing ensemble size $M$, the shifts of spaser frequency and of threshold population inversion are  significantly reduced, indicating effective restoration of spherical symmetry.

Note that the overall effect of off-resonant modes on spaser action is significantly weaker than on  single-molecule fluorescence. The calculated quantum efficiency $Q$, shown in Fig.~\ref{fig1} (see Appendix B for detail), falls below 20\%  at distances $d\sim R$, and it is even lower for smaller NPs, indicating that,  at such distances,  fluorescence is largely quenched. In contrast, spaser quenching becomes substantial only for (average) gain-NP separations well below NP size (see Fig.~\ref{fig2}), while for larger distances, spaser quenching is largely suppressed.

\section{Conclusions}

\label{sec5}

In conclusion, let us discuss the role of direct dipole coupling between gain molecules in spaser action. In small systems with $M<1000$ molecules with dipole moments aligned normally to the NP surface (maximal coupling), our numerical simulations \cite{pustovit-prb16} indicated that Coulomb shifts of molecules’ excitation energies lead to dephasing. In relatively large systems with $M$ up to $5\times 10^{4 }$ molecules with \textit{random} dipole orientations that we study here, the ensemble-averaged dipole coupling between molecules vanishes, and so the energy shifts come from the fluctuations of gain distribution, which diminish  with increasing  $M$.  Importantly, the effect of direct coupling on \textit{collective} states is much weaker than on individual QEs: for example, in the case of cooperative spontaneous emission (superradiance), the collective state that is strongly coupled to radiation (superradiant state) is \textit{unaffected} by the dipole coupling \cite{shahbazyan-prb00}, a similar behavior can be expected for stimulated emission as well. 

Let us now discuss the role of Purcell’s enhancement of spontaneous emission that is known to affect negatively the spaser threshold \cite{khurgin-apl12,khurgin-nphot14}. In fact, this effect is maximal within single-mode picture, while it is less important when higher-order \textit{dark} modes are included, which is the main topic of our paper. Indeed, Fig.~\ref{fig1} shows  quenching of  plasmon-enhanced (i.e., with the Purcell factor included) spontaneous emission by off-resonant modes, indicating that Purcell-enhanced radiative losses are much lower than \textit{overall} Ohmic losses when off-resonant modes are accounted for. Note that Fig.~\ref{fig1} illustrates the \textit{competition} between Purcell enhancement and Ohmic losses in the spontaneous emission, whereas in spaser action, these effects work \textit{in sync} against reaching the threshold. While Purcell effect is expected to alter single-mode spaser threshold, here we are interested in the quenching onset \textit{relative} to  single-mode picture, so our results  in Fig. 2 should remain intact.

Finally, we considered here a specific setup with all QEs   distributed at about equal distance to the surface of spherical metal NP, e.g., on top of dielectric shell. While within single-mode picture, the spaser threshold has been derived for arbitrary plasmonic system shapes and gain distributions \cite{shahbazyan-acsphot17}, this  configuration provides us with better control over gain coupling to off-resonant modes, and also allows better comparison to  known results for single-molecule fluorescence quenching. In a more common setup, the gain is distributed within some region comparable or exceeding  the metal volume, e.g., within the dielectric shell, implying that only a relatively small fraction of QEs, located sufficiently close to the surface, can undergo efficient energy exchange with higher-order modes decaying  rapidly outside the metal structure.Therefore, for a given gain concentration, extending the gain region size should lower the spaser threshold by suppressing quenching effects.

In summary, we studied the effect of ET between gain and off-resonant plasmon modes on spaser action. We found that the mode coupling, originating from inhomogeneity of gain distribution near the metal surface, interferes with the feedback mechanism and leads to an upward shift of spaser frequency and of population inversion threshold. We have shown that  quenching effects are restricted to a thin layer near the metal surface  and are suppressed for large gain concentration. We established a simple criterion relating spaser quenching onset to gain concentration, which we supported by numerical calculations for core-shell NP-based spasers.

\acknowledgments
This work was supported in part by the National Science Foundation under Grants No. DMR-1610427 and No. HRD-1547754.

\appendix
\section{Configurational averaging}

The angular and orientations averaging of Eq. (\ref{spasing3})  renders $m=m'$, so we set $m'=m$ and sum over $m$. In the products 
$\psi_{1m}^{(j)\ast}\psi_{l_{1}m_{1}}^{(j)}\psi_{l_{1}m_{1}}^{(k)\ast}\psi_{1m}^{(k)}$
appearing in $S_{1m,l_{1}m_{1}}S_{l_{1}m_{1}, 1m}$ only the terms with $j=k$ survive the averaging since $l_{1}>1$, thus reducing the result by factor $1/M$. The averaging over orientations is performed using the relation
\begin{equation}
\langle {\bf e}_{j}^{\alpha}{\bf e}_{j}^{\beta}{\bf e}_{j}^{\gamma}{\bf e}_{j}^{\delta}\rangle =\frac{1}{15}\left ( \delta_{\alpha\beta}\delta_{\gamma\delta}+\delta_{\alpha\gamma}\delta_{\beta\delta}+\delta_{\alpha\delta}\delta_{\gamma\beta}
    \right )
\end{equation}
and Eq. (\ref{spasing3}) takes the form
\begin{equation}
\label{spasing4}
\Omega -\left (s_{1}+\sum_{l>1}\frac{\alpha_{l}f_{l}}{\Omega-s_{l}\alpha_{l}}\right )\alpha_{1}=0,
\end{equation}
where
\begin{equation}
f_{l}=\frac{1}{45}\! \left (\frac{\mu^{2}}{\hbar}\right )^{2}\sum_{j=1}^{M} n_{j}^{2} \! 
\left [J_{1}J_{l}+
2\!  \left (J_{1}^{r}J_{l}^{r}+J_{1}^{\theta}J_{l}^{\theta}+J_{1}^{\phi}J_{l}^{\phi}\right )\right ] \!  .
\end{equation}
Here we defined
\begin{equation}
J_{l}^{\alpha}=\frac{4\pi}{2l+1}\sum_{m=-l}^{l}\nabla_{\alpha} \left [\frac{Y_{lm}(\hat{\bf r})}{r^{l+1}}\right ]\nabla_{\alpha} \left [\frac{Y_{lm}(\hat{\bf r})}{r^{l+1}}\right ],
\end{equation}
and $J_{l}=J_{l}^{r}+J_{l}^{\theta}+J_{l}^{\phi}$. Using elementary properties of spherical harmonics we find
\begin{equation}
J_{l}^{r}=\frac{(l+1)^{2}}{r^{2l+4}},
~
J_{l}^{\theta}=J_{l}^{\phi}=\frac{l(l+1)}{2r^{2l+4}}, 
~
J_{l}=\frac{(2l+1)(l+1)}{r^{2l+4}},
\end{equation}
yielding
\begin{equation}
f_{l}=\frac{2}{45}(l+1)(11l+7)\left (\frac{\mu^{2}}{\hbar}\right )^{2}\sum_{j=1}^{M} 
\frac{n_{j}^{2}}{r_{j}^{6}r_{j}^{2l+4}}.
\end{equation}
For weak dispersion of radial distribution, $r_{j}\approx r$ and weak inhomogeneity in molecular population inversion, $n_{j}\approx n=N/M$, we obtain
\begin{equation}
f_{l}=\frac{1}{5M}(11l+7)s_{1}s_{l},
\end{equation}
with $s_{l}$ given by Eq.~(\ref{sl2}), which, after being subsituted into Eq.~(\ref{spasing4}), leads to  Eq.~(\ref{spasing5}).

\section{Fluorescence quantum efficiency}

Fluorescence quantum efficiency for a single QE near metal NP has the form 
\begin{equation}
Q=\frac{\Gamma_{r}}{\Gamma_{r}+\Gamma_{nr}},
\end{equation}
where $\Gamma_{r}$ and $\Gamma_{nr}$ are, respectively, radiative and nonradiative decay rates.  For a QE oriented normally to spherical NP surface, these rates have the form \cite{nitzan-jcp81,ruppin-jcp82,pustovit-prl09,pustovit-prb10}
\begin{align}
 \Gamma _{r} =\gamma _{r}^{0} \left| 1 + \frac{2\alpha _{1}(\omega_{1})}{\left( {R + d} \right)^{3}} \right|^2\! ,
 ~
\Gamma _{nr} = \frac{3\gamma _{r}^{0} }{2k^{3}}\sum\limits_l \frac{\left( l + 1 \right)^{2} \alpha''_{l}(\omega_{1}) }{ \left( {R + d} \right)^{2l + 4}}, 
\end{align}
where $\gamma _{r}^{0}$ is the radiative decay rate for isolated QE and $k$ is the light wave vector.


%
\end{document}